\documentclass[11pt]{article}
\usepackage[margin=1in]{geometry}
\usepackage{amsmath,amssymb,amsthm,mathtools}
\usepackage{braket}
\usepackage{bm}
\usepackage{graphicx}
\usepackage{xcolor}
\usepackage[numbers,sort&compress]{natbib}
\usepackage{authblk}
\usepackage{url}

\newtheorem{theorem}{Theorem}
\newtheorem{definition}{Definition}

\newif\ifshowmajorchanges
\showmajorchangestrue

\newcommand{\changedsection}[1]{%
  \ifshowmajorchanges\section{\textcolor{black}{#1}}\else\section{#1}\fi}
\newcommand{\changedsubsection}[1]{%
  \ifshowmajorchanges\subsection{\textcolor{black}{#1}}\else\subsection{#1}\fi}
\newcommand{\changedsubsubsection}[1]{%
  \ifshowmajorchanges\subsubsection{\textcolor{black}{#1}}\else\subsubsection{#1}\fi}
\newcommand{\changedsectionstar}[1]{%
  \ifshowmajorchanges\section*{\textcolor{black}{#1}}\else\section*{#1}\fi}

\begin{document}
\sloppy

\title{Classical and quantum kernel fusion for two-sample testing}

\author[1,2]{Yu Terada\thanks{Present address: Center for AI and Data Science, Ochanomizu University, 2-1-1 Ohtsuka, Bunkyo-ku, Tokyo 112-8610, Japan}}
\author[1,2]{Yugo Ogio}
\author[1]{Ken Arai}
\author[1]{Hiroyuki Tezuka\thanks{Present address: Deloitte Tohmatsu Consulting LLC.}}
\author[1,2]{Yu Tanaka\thanks{Correspondence: yukanata@gmail.com}}
\affil[1]{Advanced Research Laboratory, Sony Group Corporation, 1-7-1 Konan, Minato-ku, Tokyo 108-0075, Japan}
\affil[2]{These authors contributed equally.}
\date{}

\maketitle

\begin{abstract}
Two-sample tests have been extensively employed in various scientific fields and machine learning to discriminate whether two sets of samples come from the same distribution or not.
Kernel-based procedures for hypothetical testing have been proposed to efficiently disentangle high-dimensional complex structures in data to obtain accurate results in a model-free way by embedding the data into the reproducing kernel Hilbert space (RKHS).
While the choice of kernels plays a crucial role for their performance, little is understood about how to choose kernel especially for small datasets.
Here we construct a hypothetical test which can be effective even for small datasets, based on the theoretical foundation of kernel-based tests using maximum mean discrepancy, which is called MMD-FUSE.
We enhance the MMD-FUSE framework by incorporating quantum kernels and propose a novel hybrid testing strategy that fuses classical and quantum kernels. 
This approach creates a powerful and adaptive test by combining the domain-specific inductive biases of classical kernels with the unique expressive power of quantum kernels. 
We evaluate our method on various synthetic and real-world clinical datasets, and our experiments reveal two key findings: 
1) With appropriate hyperparameter tuning, MMD-FUSE with quantum kernels consistently improves test power over classical counterparts, especially for small and high-dimensional data. 
2) The proposed hybrid framework demonstrates remarkable robustness, adapting to different data characteristics and achieving high test power across diverse scenarios. 
These results highlight the potential of quantum-inspired and hybrid kernel strategies to build more effective statistical tests, offering versatile tools for data analysis where sample sizes are limited.
\end{abstract}

\section{Introduction}
Statistical tests play a crucial role in extracting useful insights from observed data in a variety of scenarios such as scientific research, machine learning, industrial applications. 
In particular, two-sample tests are used for discriminating whether two sample sets were drawn from the same distribution or not, by setting the null hypothesis that they are drawn from one distribution.
For example, two-sample tests can be utilized for evaluating potential effect of new drugs and performing A/B testing with different marketing strategies \cite{aras2011tukey,larsen2024statistical}.
In the context of machine learning, two-sample tests can be used for constructing discriminator models in GAN networks \cite{li2017mmd,lopez2017revisiting,li2015generativemomentmatchingnetworks}.

Kernel-based approaches have been used for two-sample tests \cite{gretton2012kernel}, disentangling complex nonlinear relations in data to perform statistical tests reliably.
Based on maximum mean discrepancy (MMD), they address the statistical hypothetical test problem in a nonparametric way, where any specific distributions or relations on data are not assumed.
While these methods have been used extensively in various fields \cite{ozier2024kernel,bischoff2024practical,unterthiner2018towards}, it is notable that their performance may be drastically variable depending on choice of kernels.
Theoretical studies provide a principle to choose optimal kernels, using asymptotic analysis \cite{gretton2012optimal}, which succeeds in removing the arbitrariness in choice of kernels unlike the median heuristic \cite{gretton2012kernel}.
However, these results assure that one can choose such kernels only in the large sample number limit.
Hence, the use of such methods is limited in practical applications where obtaining samples is costly or collecting large datasets is infeasible.
Therefore, kernel-based two-sample testing can be interpreted as a problem of selecting the optimal kernel function to maximize test power, given the nature of the underlying data distributions.

A recent work by Schrab \textit{et al.} \cite{schrab2023mmd} develops the theory which holds even for small sample number regimes, which is called the MMD aggregated two-sample test, without depending on asymptotic theory.
However, this approach is based on multiple testing and therefore is not suitable for a large number sets of kernels.
MMD-FUSE \cite{biggs2023mmd} resolves this problem using a single statistic as a fuse combining multiple kernels.
MMD-FUSE can thus be seen as a unifying framework that formulates the kernel selection as a meta-optimization over kernel sets, potentially adapted to each data type or structure.
However, while they paved the way for finding optimal kernels, it remains unclear whether such kernels can be identified in practice and how this can be achieved.

In general, the optimal test statistic for a given task depends on the nature of the underlying data and may vary significantly across domains.
In particular, under small-sample regimes, it is crucial to tailor the test function to the structure of the data to achieve reliable statistical conclusions.
This motivates the construction of adaptive testing strategies that optimize over a collection of candidate kernels based on observed data characteristics.

In this work, we explore two complementary directions toward this goal.
First, we investigate whether quantum kernels—previously shown to be effective in small-data causal discovery \cite{terada2025quantum} can improve the performance of MMD-FUSE tests in practice.
Second, we extend this idea to hybrid kernel sets combining both classical and quantum kernels, aiming to leverage the strengths of both and enhance test power across diverse scenarios.

To investigate the potential of quantum kernels as an effective component in kernel-based two-sample testing, we first study the performance of MMD-FUSE using quantum kernels exclusively.
Quantum kernels are known to enhance performance in various kinds of tasks compared to classical kernels \cite{havlivcek2019supervised,terada2025quantum}.
In particular, the previous works showed that quantum kernels can improve performance in causal discovery for small datasets \cite{terada2025quantum}, as well as in other tasks \cite{havlivcek2019supervised,caro2022generalization}.

To evaluate the practical performance of quantum kernels in statistical hypothesis testing, we conduct experimental simulations on both synthetic and real datasets.
We first test MMD-FUSE using only quantum kernels and compare their performance with those from classically tuned kernels used in the original work \cite{biggs2023mmd}.
The results show that quantum kernels perform comparably to their classical counterparts on both synthetic and real datasets, and in some cases, they improve test power, especially under small-sample conditions.

We further explore the potential of hybrid MMD-FUSE frameworks that fuse quantum and classical kernels into a single kernel pool.
Our results indicate that such hybrid strategies retain the advantages of quantum kernels in low-sample regimes while maintaining robustness in domains where classical kernels exhibit strong inductive biases, such as clinical data.
These experiments provide empirical support for the potential benefits of both pure quantum and hybrid quantum-classical kernel constructions in kernel-based hypothesis testing.

The contributions of this work are as follows:
\begin{itemize}
  \item We introduce quantum kernels into the MMD-FUSE framework and empirically demonstrate their ability to improve test power, in particular under small-sample settings.

  \item We propose a hybrid MMD-FUSE strategy that combines classical and quantum kernels, achieving robust performance across datasets with diverse structures.

  \item Our results provide practical guidance on kernel selection and demonstrate how combining expressivity and adaptability leads to statistically powerful and domain-adaptive two-sample tests.
\end{itemize}


\long\gdef\theoreticalmethods{%
\changedsubsection{Theoretical framework}\label{sec:2}
This subsection reviews the background necessary for our proposed method, including two-sample testing, maximum mean discrepancy (MMD), permutation testing, and the MMD-FUSE~\cite{biggs2023mmd} framework, and introduces quantum kernels and the problem setting.

\subsubsection{Two-sample testing}
The two-sample testing problem is to determine whether two distributions $p$ and $q$ are equal or not. 
To be more precise, when two samples $X := (x_1, ..., x_n) \overset{\text{i.i.d.}}{\sim} p$ and $Y := (y_1, ..., y_m) \overset{\text{i.i.d.}}{\sim} q$ are given, a hypothesis test $\Delta$ is performed to evaluate the null hypothesis $H_0 : p = q$ against the alternative hypothesis $H_1 : p \neq q$. 
The hypothesis test $\Delta$ is defined as a $\{ 0, 1\}$-valued function of $Z := (X, Y) = (x_1, ..., x_n, y_1, ..., y_m)$, which rejects the null hypothesis $H_0$ if $\Delta(Z) = 1$ and fails to reject it otherwise.

It is usually designed to control the probability of a type I error at some level $\alpha \in (0,1)$, such that $\mathbb{P}_{p \times p}(\Delta(Z) = 1) \le \alpha$, while simultaneously minimizing the probability of a type II error, $\mathbb{P}_{p \times q}(\Delta(Z) = 0)$. Here, we have used the notation $\mathbb{P}_{p \times p}$ and $\mathbb{P}_{p \times q}$ to indicate that the sample $Z$ is either drawn from the null $p = q$, or the alternative $p \neq q$.
When $\mathbb{P}_{p \times q}(\Delta(Z) = 0) \le \beta \in (0,1)$, the hypothesis test $\Delta$ is said to have the power $1 - \beta$.

\subsubsection{Maximum mean discrepancy}
The Maximum Mean Discrepancy (MMD) is a kernel-based measure of distance between two distributions $p$ and $q$.
The MMD compares their mean embeddings in a reproducing kernel Hilbert space (RKHS) with kernel $k$.
Formally, if $\mathcal{H}_k$ is the RKHS associated with kernel function $k$, the MMD between distributions $p$ and $q$ is the integral
probability metric defined by
\begin{equation}
    {\rm MMD}_k (p, q) := \sup_{\substack{f \in \mathcal{H}_k, \\ \|f\|_{\mathcal{H}_k} \le 1 }} \left( \mathbb{E}_{X \sim p} [f(X)] - \mathbb{E}_{Y \sim q} [f(Y)] \right).
\end{equation}

The minimum variance unbiased estimate of ${\rm MMD}^2_k$ is given by the sum of two U-statistics~\cite{Lee2019-aa} and a
sample average as
\begin{eqnarray}
    \widehat{{\rm MMD}^2_k}(Z) 
    &:=& 
    \frac{1}{n(n-1)} \sum_{(i,i' \in [n]_2)} k(X_i, X_{i'}) \nonumber \\
    && + \frac{1}{m(m-1)} \sum_{(j,j' \in [m]_2)} k(Y_j, Y_{j'}) \nonumber \\
    && - \frac{2}{nm} \sum_{i \in [n]} \sum_{j \in [m]} k(X_i, Y_j),
\end{eqnarray}
where we introduced the notation $[n]_2 = \{ (i,i') \in [n]^2 : i \neq i' \}$ for the set of all pairs of distinct
indices in $[n] = \{1, ... , n \}$.

\subsubsection{Permutation tests}
Permutation testing is a non-parametric method that constructs a null distribution by randomly shuffling the labels of the sample, without assuming any specific underlying distribution, and assesses the significance of the test statistic accordingly. 
The tests use permutations of the data $Z$ to approximate the null distribution. 
Let $S_{n+m}$ denote the permutation (or symmetric) group on $[n+m]$, and let $\sigma \in S_{n+m}$ be a permutation, {\it i.e.}, a bijective map from $[n+m]$ to $[n+m]$. 
Furthermore, we denote the label permutation of the data $Z$ as $\sigma Z = (Z_{\sigma(1)}, ..., Z_{\sigma(n+m)})$.
Then, it is clearly that the permuted data $\sigma Z$ simulates the null distribution.

We can use permutations to construct an approximate cumulative distribution function (CDF) of the test statistic. 
By choosing an appropriate quantile as the threshold, we can then perform the test.
The quantile operator for a finite set $\{ f(a) \in \mathbb{R} : a \in \mathcal{A} \}$ is defined by
\begin{equation}
    \mathop{{\rm quantile}}_{q, a \in \mathcal{A}} f(a) := \inf \left\{ r \in \mathbb{R} : \frac{1}{|\mathcal{A}|} \sum_{a \in \mathcal{A}} \mathbf{1} \{ f(a) \le r \} \ge q \right\}.
\end{equation}

The following theorem is used to construct a practical test.
\begin{theorem}[Theorem 2 in Ref.~\cite{hemerik2018exact}] \label{thm:perm_test}
    Let $G$ be a vector of elements from $S_n$, $G = (\sigma_1, ..., \sigma_{B}, \sigma_{B+1})$, with $\sigma_{B+1} = id$ (the identity permutation) for any $B \ge 1$. The elements $\sigma_1, ..., \sigma_{B}$ are drawn uniformly from $S_n$ either i.i.d. or without replacement (which includes the possibility of $G = S_n$). If $\tau(Z)$ is a statistic of $Z$ and $Z \overset{d}{=} \sigma Z$ for all $\sigma \in S_n$ under the null then
    \begin{equation}
        \mathbb{P}_{p \times p, G} \left( \tau(Z) \ge \mathop{{\rm quantile}}_{1-\alpha, \sigma \in S_n} \tau(\sigma Z) \right) \le \alpha. 
    \end{equation}
\end{theorem}
According to this theorem, constructing a permutation test reduces to designing an appropriate test statistic. 
This result holds exactly for any number of randomized permutations $B \ge 1$, allowing for a straightforward construction of an exact and computationally efficient test.

In practice, a test statistic $\tau(Z)$ is chosen (e.g., the MMD or FUSE-1 statistic introduced later), and the $(1-\alpha)$ quantile of its permuted values is used as a threshold.
We denote this threshold by $\tau_\alpha$, and define the hypothesis test $\Delta(Z)$ as rejecting $H_0$ if $\tau(Z) \ge \tau_\alpha$.

\subsubsection{MMD-FUSE}
Reference~\cite{biggs2023mmd} proposed a method called MMD-FUSE for constructing a permutation test by combining $\widehat{{\rm MMD}}_k^2$ values calculated under different kernels $k \in \mathcal{K}$. 
The key idea is to define a test statistic as the soft maximum (log-sum-exp) over MMD values, which stabilizes the effect of kernel scaling and prevents overfitting that could arise when using the maximum or naive summation.

Specifically, MMD-FUSE assumes that each kernel $k$ is drawn from a prior distribution $\pi$ over a finite kernel set $\mathcal{K}$ with $|\mathcal{K}| = r$, typically taken to be the uniform distribution.
Then, the test statistic called FUSE-1 is defined as follows:
\begin{definition}[Definition 1 in Ref.~\cite{biggs2023mmd}]
    We define the un-normalized test statistic with parameter $\lambda > 0$ as
    \begin{equation}
        \widehat{{\rm FUSE}}_1 (Z) := \frac{1}{\lambda} \log \left( \mathbb{E}_{k \sim \pi} \left[ \exp\left(\lambda \widehat{{\rm MMD}}_k^2 (X, Y) \right)\right] \right). \label{eq:fuse1}
    \end{equation}
\end{definition}

Since the prior $\pi$ is supported on a finite set $\mathcal{K}$, the above expectation reduces to a simple average, and the test statistic becomes a log-sum-exp over the empirical MMD estimates.

The soft maximum plays a crucial role in avoiding overfitting.
If one were to simply take the maximum of $\widehat{{\rm MMD}}_k^2$ across $k \in \mathcal{K}$, the statistic would increase as more kernels are added, regardless of their relevance, leading to power saturation and potential bias.
Moreover, the variance and scale of MMD can vary significantly depending on the kernel bandwidth.
The soft aggregation mitigates this by weighing the contributions more smoothly.

Having defined the test statistic as in Eq.~(\ref{eq:fuse1}), we can now perform a two-sample test by applying it within the framework of Theorem~\ref{thm:perm_test}.
In this study, we investigate how the choice of the kernel set $\mathcal{K}$—including quantum, classical, and hybrid constructions—affects the test power of MMD-FUSE.

\subsubsection{Quantum kernels}
Quantum kernels embed data into the RKHS associated with quantum models, which may enhance computational performance in machine learning and statistical tasks~\cite{schuld2021supervised,terada2025quantum}.  
Here, quantum kernels are defined by quantum circuits that have inputs and their states: the kernel matrices are defined as
\begin{equation}
k_{\mathrm{Q}}(\mathbf{x},\mathbf{x}') = \mathrm{Tr}[\rho(\mathbf{x}) \rho(\mathbf{x}')], \label{eq:q_kernel}
\end{equation}
where $\mathbf{x}$ and $\mathbf{x}'$ represent data and $\rho(\mathbf{x})$ denotes the quantum state generated by the quantum circuits.  
In this work, we employ the hyperparameter only for scaling in preprocessing and use fidelity as the observable parameter.  

In the simplest case, a single-qubit quantum state is initialized in $\ket{+}$ and rotated by the $y$-axis unitary gate:
\begin{equation}
\ket{\psi(x)} := R_y(x) \ket{+}.
\end{equation}
The resulting quantum kernel is given by the fidelity between states,
\begin{equation}
k_{\mathrm{Q}}(x, x') = |\braket{\psi(x)}{\psi(x')}|^2 = \cos^2\left( \frac{x - x'}{2} \right). 
\end{equation}

Although this represents a minimal case, increasing the number of qubits and the circuit depth allows for the construction of quantum kernels with higher expressive capacity.  
Indeed, the expressivity of a quantum kernel, which reflects the richness of its associated reproducing kernel Hilbert space (RKHS), strongly depends on the structure of the quantum feature map and the data encoding scheme~\cite{schuld2020effect}.  
Carefully chosen quantum embeddings may even yield kernels that are provably hard to compute classically, thereby providing a route toward quantum advantage in supervised and unsupervised learning.

\subsubsection{Problem setting}
Given a sample $Z = (X, Y)$ from two distributions $p$ and $q$, our goal is to construct a two-sample test $\Delta(Z)$ based on the MMD-FUSE statistic that achieves high power while controlling the type I error under the permutation framework.

To this end, we aim to choose the kernel set $\mathcal{K}$ (and thereby the prior $\pi$) so as to maximize the test power of the FUSE-1 statistic.
Formally, let $\widehat{\mathrm{FUSE}}_1^{\mathcal{K}}(Z)$ denote the test statistic using kernel set $\mathcal{K}$, and define the test via the permutation threshold $\tau_{\alpha}$ as
\begin{equation}
\Delta(Z) := \mathbf{1} \left\{ \widehat{\mathrm{FUSE}}_1^{\mathcal{K}}(Z) \ge \tau_{\alpha} \right\}.
\end{equation}
Our objective is to select $\mathcal{K}$ to maximize the power $1 - \mathbb{P}_{p \times q}(\Delta(Z) = 0)$, for fixed $\alpha$ and data $Z$.
We explore both quantum-only and hybrid quantum-classical kernel constructions to achieve this goal under various data conditions.
}

\changedsection{Results}\label{sec:3}
\changedsubsection{Quantum MMD-FUSE}
In this section, we propose an extension of the MMD-FUSE framework that incorporates quantum kernels into the kernel pool.
Motivated by recent studies showing the expressive capacity of quantum kernels in small-sample regimes \cite{terada2025quantum}, we investigate whether this integration improves the test power of kernel-based two-sample testing.
We also evaluate the behavior of this approach under various data conditions, including synthetic and real datasets, and compare it to classical kernel collections.

\long\gdef\quantummodelmethods{%
\changedsubsection{Quantum-kernel MMD-FUSE}
We extend the MMD-FUSE framework by replacing the classical kernel set with a collection of quantum kernels. 
Let $\mathcal{K}_{\mathrm{Q}} = \{ k^{(1)}_{\mathrm{Q}}, \ldots, k^{(r)}_{\mathrm{Q}} \}$ denote a finite set of quantum kernels, where each $k^{(i)}_{\mathrm{Q}}$ is defined by a quantum feature map and corresponding fidelity-based kernel as described in Eq.~(\ref{eq:q_kernel}). 
Given this kernel set, we define the FUSE-1 test statistic using the log-sum-exp formulation:
\begin{equation}
\widehat{\mathrm{FUSE}}_1^{\mathcal{K}_{\mathrm{Q}}}(Z) := \frac{1}{\lambda} \log \left( \frac{1}{r} \sum_{k \in \mathcal{K}_{\mathrm{Q}}} \exp\left(\lambda \widehat{\mathrm{MMD}}_k^2(Z)\right) \right),
\end{equation}
where $\lambda > 0$ is a temperature parameter and $\widehat{\mathrm{MMD}}_k^2(Z)$ is the unbiased MMD estimate for kernel $k$.
The corresponding two-sample test is then defined via the permutation threshold $\tau_\alpha$ as $\Delta(Z) := \mathbf{1}\{ \widehat{\mathrm{FUSE}}_1^{\mathcal{K}_{\mathrm{Q}}}(Z) \ge \tau_\alpha \}$.

In our implementation, we construct the kernel set $\mathcal{K}_{\mathrm{Q}}$ by varying a scaling parameter applied to the input features prior to quantum state encoding. 
This scaling controls the spread of the data in Hilbert space and significantly influences the behavior of the quantum kernels. 
Following prior work \cite{terada2025quantum}, we generate multiple quantum kernels by sweeping this scaling parameter over a logarithmic grid (e.g., $10^{-3}$ to $10^0$), ensuring diversity in the kernel set.
Each kernel corresponds to a quantum feature map that encodes a data point $\bm{x}$ into a quantum state $\rho(\bm{x})$, with the kernel function defined as $k_{\mathrm{Q}}(\bm{x}, \bm{x}') = \mathrm{Tr}[\rho(\bm{x})\rho(\bm{x}')]$.
The final kernel pool $\mathcal{K}_{\mathrm{Q}}$ thus consists of fidelity-based kernels differentiated by their input scaling.
}

\long\gdef\quantumexperimentmethods{%
\changedsubsection{Datasets and evaluation protocol}\label{subsec:3_2}
To evaluate the effectiveness of MMD-FUSE with quantum kernels, we conduct two-sample tests using one synthetic and two real-world datasets.

The synthetic dataset that we used here consists of samples generated from multivariate Gaussian distributions (or log-normal ) with a mean shift.
Each sample includes examples of $D$-dimensional vectors in $\mathbb{R}^D$, where $D = 2$ or $D = 6$ depending on the experimental setting.
Specifically, we generate two samples $X^{(1)} = \{ \bm{x}^{(1)}_i \in \mathbb{R}^D \}_{i=1}^M$ and $Y^{(1)} = \{ \bm{y}^{(1)}_i \in \mathbb{R}^D \}_{i=1}^M$ following
\begin{align}
x^{(1)}_{ij} &\sim \mathcal{N}(0, 1), \label{eq:data1_1} \\
y^{(1)}_{ij} &\sim \mathcal{N}(d, 1), \quad \text{for } j = 1, \dots, M, \label{eq:data1_2}
\end{align}
where $d$ controls the mean difference between the two distributions along each coordinate.
In our experiments, we considered both low-dimensional ($D = 2$) and higher-dimensional ($D = 6$) cases to investigate the effect of input dimensionality on test performance.
In both cases, the two distributions differ only in the means of their distributions, and the success of the two-sample test is monitored by its ability to reject the null hypothesis $H_0 : p = q$ with a fixed value of level of significance.

The first real-world dataset is related to clinical heart disease \cite{ahmad2017survival}.
It contains 12 clinical variables, including measurements of ejection fraction and serum creatinine, which have been identified as relevant indicators of patient outcomes in prior studies \cite{chicco2020machine}.
The dataset is divided into two groups: individuals who experienced death events and those who did not.
Let $X^{(2)}$ denote the samples from the non-death group ($n = 203$) and $Y^{(2)}$ the samples from the death group ($n = 96$).
In our experiments, we consider two settings: (i) a low-dimensional case using only the two variables (ejection fraction and serum creatinine), and (ii) a high-dimensional case using all 12 variables.
Each sample is therefore either a 2D or 12D vector depending on the setting.

The second real-world dataset is associated with observations on breast cancer \cite{street1993nuclear}.
The dataset includes 30 feature variables and 1 label.
The label indicates whether the tumor is benign or malignant.
The number of benign examples is 357 while that of malignant is 212.
We use two crucial features associated on concavity of tumors in our experiment.

In all experiments, we vary the sample size to evaluate how test power changes under limited data availability.
The number of permutation trials for estimating the null distribution is set to $B = 2000$, and the significance level is fixed as $\alpha = 0.05$.
We compare the test power—defined as the proportion of correctly rejecting the null hypothesis when $p \neq q$—between MMD-FUSE with quantum kernels and with classically tuned Gaussian kernels.
We also evaluate the actual significance level (empirical type I error) using data sampled from the same distribution, which is constructed by mixing the data in some cases, to assess the statistical validity of each method.
}

\begin{figure*}[htbp]
  \centering
    \includegraphics[width=0.95\linewidth]{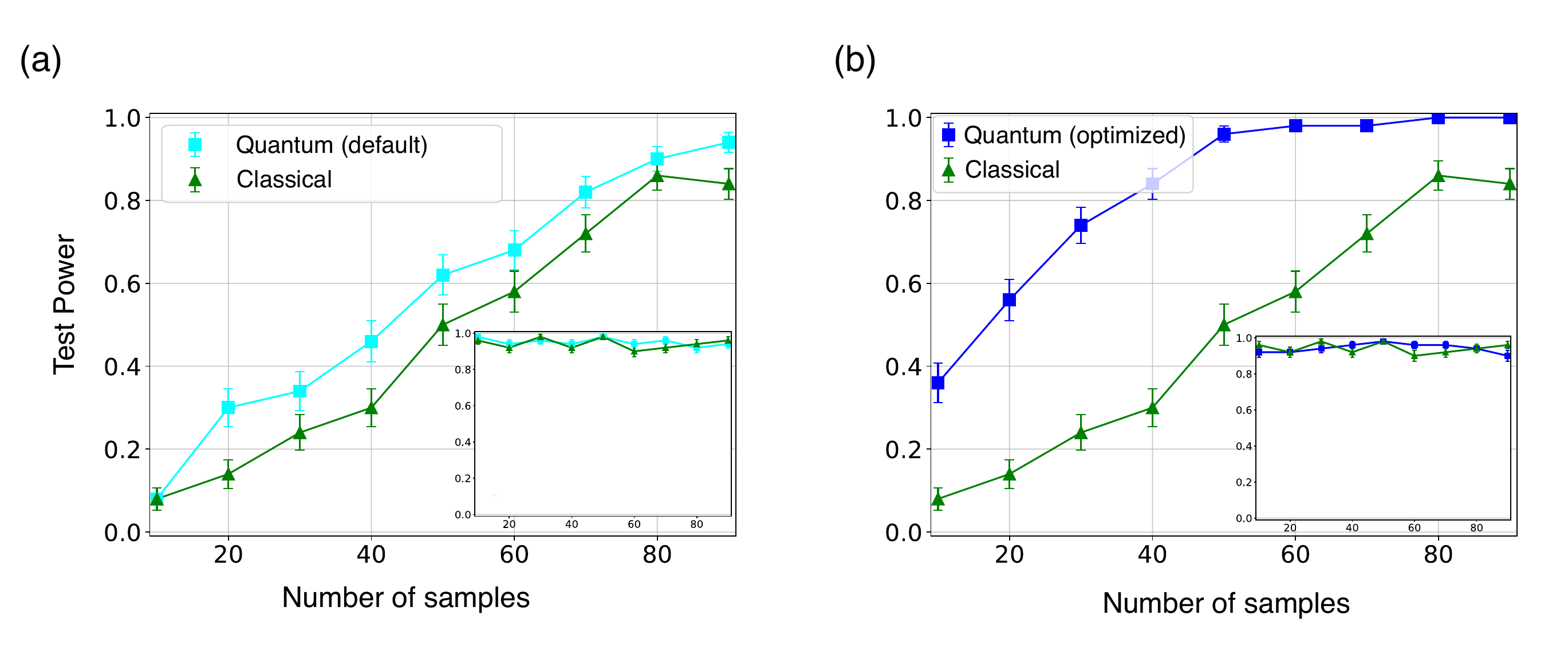}
  \caption{
    Estimates of test power for MMD-FUSE with quantum and classical kernels for synthetic Gaussian data.
    Each test was conducted by randomly selecting $10, 20, \ldots, 90$ samples from $M = 500$ samples of $X^{(1)}$ and $Y^{(1)}$, repeated $50$ times. 
    Quantum kernels use fidelity-based functions, while classical kernels use Gaussian the kernels with $8$ different bandwidths. 
    The bandwidths were determined by coverage-based scaling, and all kernels are equally weighted with weight. 
    The level of significance was set as $\alpha=0.05$, the error bars denote the standard errors across 50 simulations, and the insets show true negative rate.
    The insets show true negative rate versus sample size for shuffled distributions.
    (a) Test power versus sample size using default scaling parameters for quantum kernels with those using the classical kernels. 
    (b) Test power of quantum kernels with the scaling parameters optimized in the previous work \cite{terada2025quantum}.
    }
  \label{twofigs:results_mmd_fuse_with_q_kernels_for_data1}
\end{figure*}

\changedsubsubsection{Quantum-kernel performance}
First, the comparison of estimated test powers obtained from the two-sample tests based on MMD-FUSE with quantum kernels to that with classical kernels are shown in Fig.~\ref{twofigs:results_mmd_fuse_with_q_kernels_for_data1}, where we evaluate performance on the synthetic Gaussian dataset with $D=2$ and mean shift $d=0.5$.
Figure~\ref{twofigs:results_mmd_fuse_with_q_kernels_for_data1} (a) shows the results using the fixed scaling parameters for quantum kernels, while Figure~(b) shows the results after optimizing the scaling parameters that are obtained in \cite{terada2025quantum}.
In both cases, the test power increases with the sample size, but the optimized scaling in (b) leads to consistently higher performance across the sample range.
This confirms that appropriate scaling of quantum kernels can significantly enhance test power, especially in the small- to moderate-sample regime, and allows quantum kernels to perform competitively with classically tuned kernels.

\begin{figure*}[htbp]
  \centering
    \includegraphics[width=0.95\linewidth]{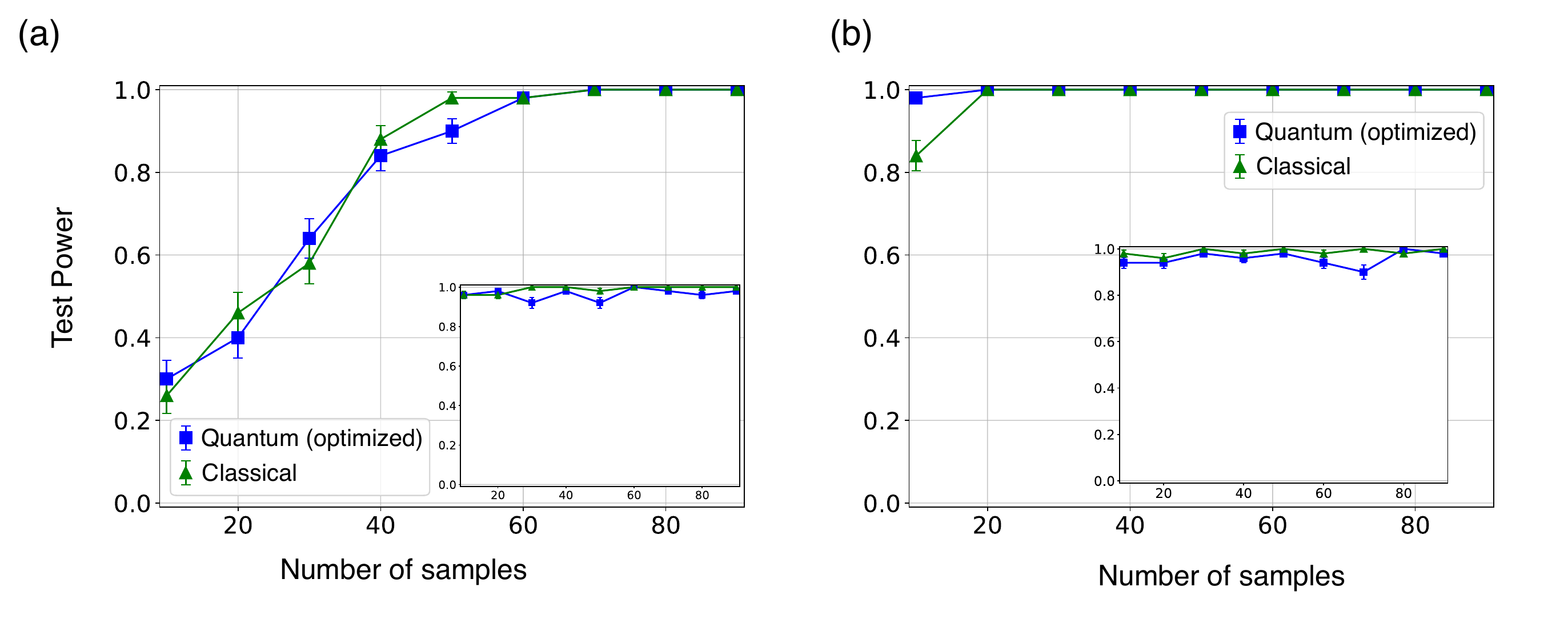}
  \caption{
    Estimates of test power for MMD-FUSE with quantum and classical kernels for the two real-world datasets on heart disease and breast cancer data. 
    Tests were conducted by randomly selecting $10, 20, \ldots, 90$ samples from two groups divided depending on primary dichotomous variables.
    Each sample consist of a 2 dimensional vector.
    Quantum kernels are equipped with the scaling parameters optimized in the previous work \cite{terada2025quantum}.
    The error bars denote the standard errors across 50 independent simulations and the insets show true negative rate versus sample size for shuffled distributions.
    (a) Test power versus sample size using quantum and classical kernels. Clinical dataset on heart disease with 203 patients with survival and 96 with death events were used. Their feature variables denote ejection fraction and serum creatinine. 
    (b) Test power on clinical dataset on breast cancer data with 357 benign and 212 malignant tumors.
    }
    \label{twofigs:results_mmd_fuse_with_q_kernels_for_data2}
\end{figure*}

After we confirmed the performance gain observed for the Gaussian data in Fig.~\ref{twofigs:results_mmd_fuse_with_q_kernels_for_data1}, we applied quantum kernels with optimized scaling parameters to two real-world datasets on heart disease and breast cancer.
Figure~\ref{twofigs:results_mmd_fuse_with_q_kernels_for_data2} (a) and (b) show the test power achieved on these datasets.
In both cases, the optimized quantum kernels provide accurate results, which are compatible with those by classical kernels.
They suggest that the scaling optimization strategy developed for synthetic data can generalize effectively to real-world applications across different domains.

\begin{figure*}[htbp]
  \centering
    \includegraphics[width=0.95\linewidth]{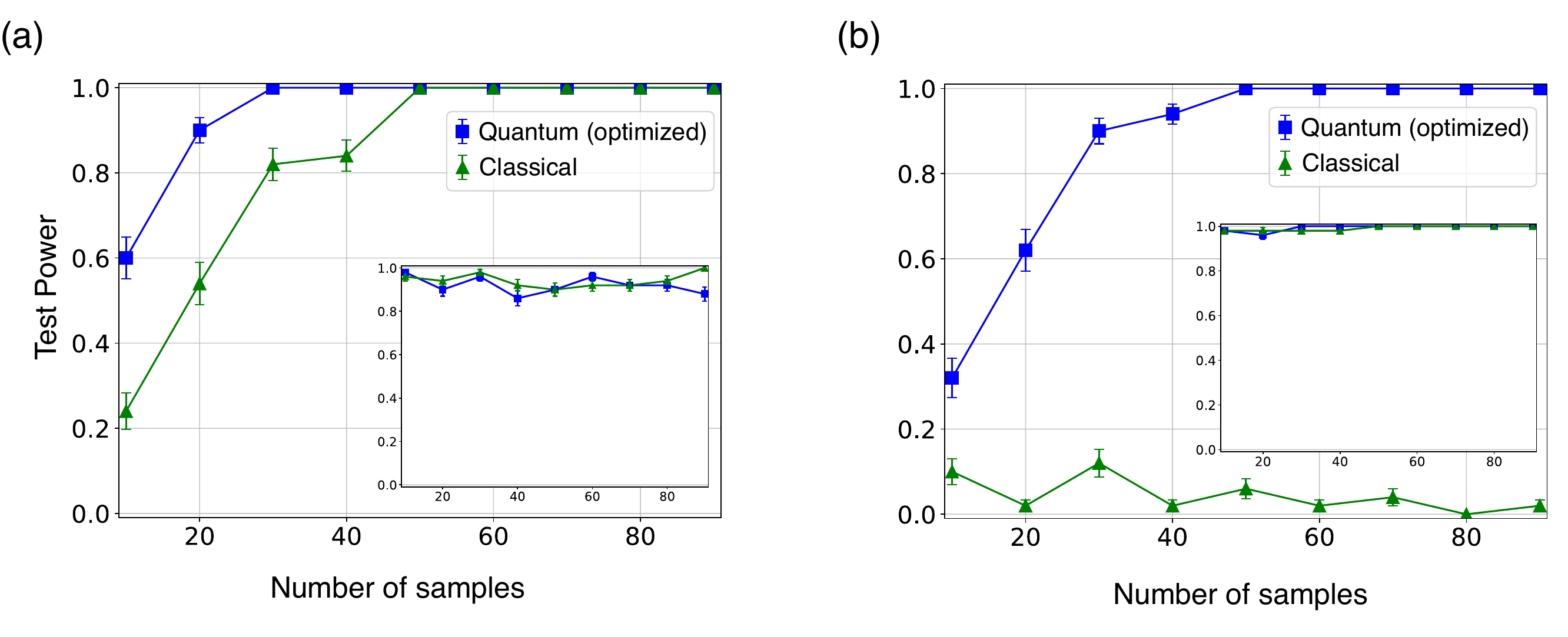}
  \caption{
    Estimates of test power for MMD-FUSE with quantum and classical kernels on high-dimensional datasets. 
    The datasets were generated following the same procedures as in Figs. \ref{twofigs:results_mmd_fuse_with_q_kernels_for_data1} and \ref{twofigs:results_mmd_fuse_with_q_kernels_for_data2} (a), but the dimensions of the datasets are higher than them.
    The error bars denote the standard errors across 50 independent simulations and the insets show true negative rate versus sample size for shuffled distributions.
     (a) Synthetic Gaussian dataset with $D = 6$.
     (b) Heart disease dataset with $D = 12$, where all 12 observed clinical variables were used. 
    }
    \label{twofigs:results_mmd_fuse_with_q_kernels_for_data3}
\end{figure*}

To further evaluate the effectiveness of quantum kernels in high-dimensional settings, we applied MMD-FUSE to the synthetic Gaussian dataset with $D=6$ and heart disease dataset with $D=12$, using all available features in each dataset.
Figure~\ref{twofigs:results_mmd_fuse_with_q_kernels_for_data3} (a) and (b) show the estimated test powers for these cases.
In both datasets, the quantum kernels consistently outperform classical kernels, especially when the sample size is small.
These results demonstrate that MMD-FUSE with quantum kernels retains its discriminative power even as the input dimensionality increases, highlighting their practical value in high-dimensional real applications.

Across all experimental conditions ranging from low-dimensional synthetic data to high-dimensional real-world datasets, the integration of quantum kernels into the MMD-FUSE framework consistently improved test power compared to that with the use of classical kernels, especially under limited sample sizes.
This advantage was observed not only in synthetic settings, where scaling parameters could be tuned explicitly, but also in multiple real biomedical datasets with complicated statistical structures.
Our results indicate that quantum kernels provide a flexible and expressive hypothesis space that adapts well across domains and dimensionalities.

\changedsubsection{Hybrid MMD-FUSE}\label{sec:4}
While quantum kernels have demonstrated promising performance in small-sample scenarios for both synthetic and real data, classical kernels often exhibit strong inductive biases that align well with certain data.
To combine the complementary strengths of both kernel types, we propose a hybrid MMD-FUSE framework that unifies classical and quantum kernels into a single kernel pool.
This section presents the formulation of the hybrid approach, followed by its empirical evaluation on both synthetic and real-world datasets, and demonstrates its robustness across diverse settings.

\long\gdef\hybridmodelmethods{%
\changedsubsection{Hybrid kernel construction}
The hybrid framework extends the original method by combining classical and quantum kernels within a kernel set.
Let $\mathcal{K}_{\mathrm{H}} = \mathcal{K}_{\mathrm{C}} \cup \mathcal{K}_{\mathrm{Q}}$, where $\mathcal{K}_{\mathrm{C}}$ denotes a set of classical kernels such as Gaussian and Laplace with varying bandwidths, and $\mathcal{K}_{\mathrm{Q}}$ denotes a set of quantum kernels generated by varying hyperparameters such as a scaling parameter.
The rationale behind this hybridization is that classical kernels often exhibit domain-specific inductive biases (e.g., smoothness, locality), while quantum kernels offer a richer hypothesis space and improved performance in small-sample regimes.
By fusing these two types of kernels, we aim to improve test power across a broader range of data distributions and sample sizes.

In our implementation, we construct $\mathcal{K}_{\mathrm{C}}$ using Gaussian kernels with ten bandwidths each, and $\mathcal{K}_{\mathrm{Q}}$ using quantum kernels with five logarithmically spaced scaling parameters, resulting in a hybrid kernel pool of size $r = 25$.
To ensure uniform contribution across kernels, we adopt an equal weighting scheme in the FUSE-1 aggregation, effectively setting the prior $\pi$ over $\mathcal{K}_{\mathrm{H}}$ to the uniform distribution.
While the inclusion of quantum kernels increases the computational cost due to quantum state preparation and fidelity computation, the permutation-based testing framework remains unchanged.
This hybrid approach thus preserves the simplicity of the original MMD-FUSE design while expanding its representational capacity through kernel diversity.
}

\long\gdef\hybridexperimentmethods{%
\changedsubsection{Hybrid evaluation protocol}
To evaluate the effectiveness of the hybrid MMD-FUSE framework, we perform comparative two-sample tests using three different kernel sets:  
(i) classical kernels only ($\mathcal{K}_{\mathrm{C}}$),  
(ii) quantum kernels only ($\mathcal{K}_{\mathrm{Q}}$), and  
(iii) hybrid kernels ($\mathcal{K}_{\mathrm{H}} = \mathcal{K}_{\mathrm{C}} \cup \mathcal{K}_{\mathrm{Q}}$).

We apply the above kernel sets to three datasets: the synthetic Gaussian data, the real-world data on heart disease, and the synthetic dataset consisting of 2-dimensional vectors sampled from two log-normal distributions with shifted means.

In particular, the last dataset is generated by sampling variables for each dimension independently as follows:
\begin{align}
x^{(4)}_{i1}, x^{(4)}_{i2} &\sim \mathrm{LogNormal}(0, 1), \\
y^{(4)}_{i1}, y^{(4)}_{i2} &\sim \mathrm{LogNormal}(d, 1),
\end{align}
where $d$ denotes the amount of shift in the logarithmic mean. This setup introduces non-Gaussianity and skewness in the marginal distributions, providing a more challenging scenario for kernel-based methods.

All kernel collections were constructed using the same number of bandwidths or scaling factors as in the previous sections, and the test power is evaluated under varying sample sizes.
This experimental design allows us to directly compare the influence of kernel diversity on test power and to assess whether the hybridization improves robustness across a broader range of data characteristics, including symmetric versus skewed and synthetic versus real-world distributions.
}

\begin{figure*}[htbp]
  \centering
    \includegraphics[width=0.95\linewidth]{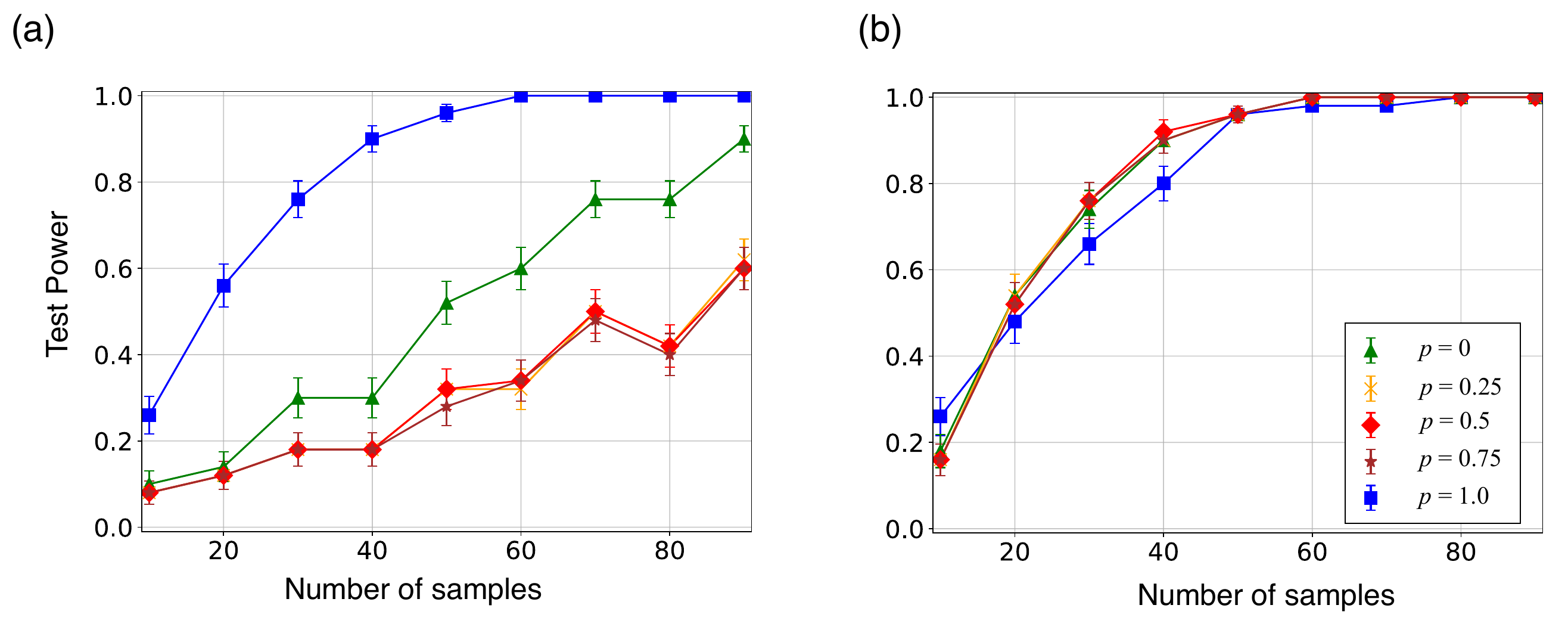}
  \caption{
    Estimates of test power for hybrid MMD-FUSE employing both classical and quantum kernels on synthetic and real-world datasets, under priors with different weights.
    The fraction $p$ for weights varies from 0 to 1, where the former corresponds to the purely classical case and the latter to the purely quantum one.
    The datasets were generated following the same procedures as in Figs. \ref{twofigs:results_mmd_fuse_with_q_kernels_for_data1} and \ref{twofigs:results_mmd_fuse_with_q_kernels_for_data2} (a), but the dimensions of the datasets are higher than them.
    The error bars denote the standard errors across 50 independent simulations.
    (a) Gaussian dataset that are the same as in Figs. \ref{twofigs:results_mmd_fuse_with_q_kernels_for_data1}, 
    (b) Clinical dataset the same as in \ref{twofigs:results_mmd_fuse_with_q_kernels_for_data2} (a).
    }
    \label{twofigs:results_hybrid_mmd_fuse_for_data}
\end{figure*}

\changedsubsubsection{Hybrid-kernel performance}
The performance of hybrid MMD-FUSE on the Gaussian and heart-disease data is summarized in Fig.~\ref{twofigs:results_hybrid_mmd_fuse_for_data}, with and without prior (kernel weight) adjustment.
The continuous parameter $p$ controls the degree of hybridization, in which the two edges $p=0$ and $p=1$ denote pure classical and quantum kernel sets, respectively.

In both datasets, when the prior over the hybrid kernel set is uniform with $p=0.5$ (i.e., no adjustment), the test power closely tracks that of classical kernels.
This indicates that, without proper weighting, the FUSE statistic is dominated by classical kernels (possibly due to their larger output scale), leading to limited benefit from the quantum components.
However, once the prior is adjusted to give more balanced emphasis (or up-weight quantum kernels), the test power can improve drastically.
This demonstrates that careful prior tuning enables the hybrid method to effectively combine classical inductive bias with quantum expressivity.

\begin{figure*}[htbp]
  \centering
    \includegraphics[width=0.49\linewidth]{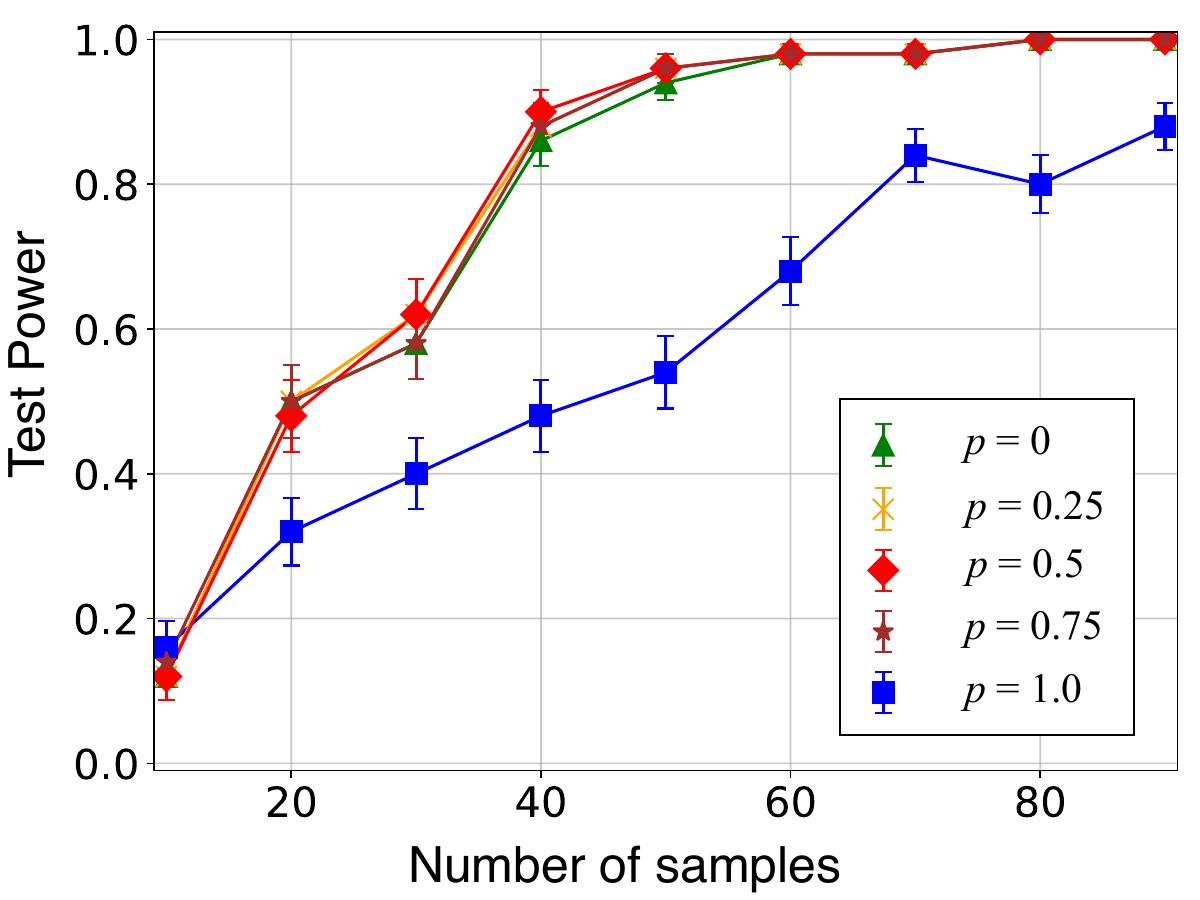}
  \caption{
    Estimates of test power for hybrid MMD-FUSE with classical and quantum kernels for the synthetic log-normal data under priors with different weights. The variables drawn from the log-normal distributions are 2D vectors ($D = 2$).
    The parameter $p$ controls of the weights for classical and quantum kernels as in Fig. \ref{twofigs:results_hybrid_mmd_fuse_for_data}.
    The error bars denote the standard errors across 50 independent simulations.
    }
\label{twofigs:results_hybrid_mmd_fuse_for_data4}
\end{figure*}

To investigate whether hybrid MMD-FUSE can still be effective when classical kernels are better suited to the data, that is, whether we can in general exploit the advantages of both classical and quantum kernels by using the hybrid method, we evaluate its performance on the synthetic dataset constructed from log-normal distributions.
Due to the skewed nature of the marginals, classical kernels tend to show better performance, especially under limited data.
Figure~\ref{twofigs:results_hybrid_mmd_fuse_for_data4} shows the test power of hybrid MMD-FUSE on this dataset with and without prior (kernel weight) adjustment.
In contrast to the previous cases, the pure classical kernel sets shows better performance than the pure quantum kernel sets.
Hybrid MMD-FUSE with well-tuned combination parameters succeeded in reproducing such high accuracy by the classical kernels.

These results highlight the flexibility of the hybrid MMD-FUSE framework in adapting to various data scenarios. 
When quantum kernels are more accurate, as in small-sample or high-dimensional settings in the Gaussian and real datasets, the hybrid method can emphasize them appropriately; when classical kernels are better suited, as in skewed distributions, the method can shift its focus accordingly.
This adaptability is made possible through simple prior (kernel weight) tuning.
Thus, by defaulting to a hybrid kernel pool, one can maintain robust test power without needing to preselect the best kernel type for each problem. 
This makes hybrid MMD-FUSE a practical and general-purpose strategy for two-sample testing across diverse data types.

\changedsection{Discussion}\label{sec:5}
In this paper, we developed a kernel selection strategy within the MMD-FUSE framework to maximize two-sample test power in a data-adaptive manner.
As one step, we proposed incorporating quantum kernels into MMD-FUSE and evaluated their effectiveness through a series of two-sample tests using both synthetic and real-world datasets.

First, we confirmed that in low-sample regimes, the test power of MMD-FUSE declines significantly when using classical kernels, whereas quantum kernels with properly tuned scaling parameters can substantially improve test power.
This was especially highlighted for the Gaussian and high-dimensional cases.
In the clinical data, we observed that quantum kernels perform competitively or better compared to classical kernels.
Furthermore, we introduced a hybrid MMD-FUSE framework that combines classical and quantum kernels, and showed that with potentially adjusted priors, it consistently could achieve or exceed the performance of individual kernel.

These results demonstrate that quantum kernels enhance the flexibility and effectiveness of kernel-based hypothesis testing and that hybrid kernel collections offer a robust and adaptive strategy across various settings including small-sample, high-dimensional, and non-Gaussian distributions.

While we demonstrated the empirical benefits of tuning scaling parameters and prior weights, we did not propose a systematic method for optimizing kernel selection or weighting.
An important direction for future work is to develop data-driven approaches for optimizing both the kernel pool and its associated hyperparameters, enabling automated and interpretable kernel design within the MMD-FUSE framework that works even for small data cases.

\changedsection{Methods}\label{sec:methods}
\theoreticalmethods
\quantummodelmethods
\quantumexperimentmethods
\hybridmodelmethods
\hybridexperimentmethods

\changedsectionstar{Acknowledgments}
We are grateful to Prof. Kenji Fukumizu for drawing our attention to related literature \cite{biggs2023mmd} that shaped this project.

\bibliographystyle{naturemag}
\bibliography{sample}

\end{document}